# The links between magnetic fields and filamentary clouds III: field regulated mass cumulative functions

Law, C.Y.[1]★ Li, H. -B.,[1]† Cao,Zhuo.,[1] and Ng, C. -Y.,[2]

[1]*Department of Physics, The Chinese University of Hong Kong, Ma Liu Shui, Shatin, NT, Hong Kong SAR*
[3]*Department of Physics, The University of Hong Kong, Pokfulam Road, Hong Kong SAR*

**ABSTRACT**
During the past decade the dynamical importance of magnetic fields in molecular clouds has been increasingly recognized, as observational evidence has accumulated. However, how a magnetic field affect star formation is still unclear. Typical star formation models still treat a magnetic fields as an isotropic pressure, ignoring the fundamental property of dynamically important magnetic fields: their direction. This study builds on our previous work which demonstrated how the mean magnetic field orientation relative to the global cloud elongation can affect cloud fragmentation. After the linear mass distribution reported earlier, we show here that the mass cumulative function (MCF) of a cloud is also regulated by the field orientation. A cloud elongated closer to the field direction tends to have a shallower MCF, in other words, a higher portion of the gas in high density. The evidence is consistent with our understanding of bimodal star formation efficiency discovered earlier, which is also correlated with the field orientations.

**Key words:** stars: formation – ISM: clouds – ISM: magnetic fields

## 1 INTRODUCTION

We have conducted a series of studies on the connections between molecular cloud fragmentation and cloud-field offset (i.e., the misalignment between the ambient magnetic field direction and the cloud elongation) (Li et al. 2013, 2017; Law et al. 2019). The series was initiated with the discovery of bimodal cloud-field offsets: Li et al. (2013) (hereafter paper I) studied 13 nearby Gould Belt molecular clouds and revealed that the long axes of a cloud tend to align either perpendicularly (large cloud-field offset) or in parallel (small cloud-field offset) to the mean directions of the ambient magnetic field. The molecular cloud long axes were derived from archival extinction maps (Dobashi 2011), and the magnetic field directions were inferred from archival starlight polarimetry data (Heiles 2000). A possible effect of the cloud-field offset on cloud fragmentation is due to the magnetic flux. For a cylindrical cloud, the largest possible magnetic flux occurs when the field is perpendicular to the long axis of the cylinder. The effect should be reflected in the mass distributions and star formation properties of the molecular clouds. Using the same dust extinction maps (Dobashi 2011), Law et al. (2019)(hereafter paper II) observed more uniform linear mass profiles for molecular clouds with large cloud-field offsets. Moreover, based on archival star formation rate data (Heiderman et al. 2010; Lada et al. 2010), Li et al. (2017) (hereafter Li17) observed lower star formation efficiency for molecular clouds with large cloud-field offsets. The present study, continuing the series, considers the effect of the cloud-field offset on the mass portion of gas in high density.

　　The paper is organized as follows. In Section 2, we first briefly describe the data and the selected molecular clouds. We then present our method for evaluating the slope of the MCF. We present the results and the statistical analyses in Section 3. A discussion of our findings follows in Section 4. The conclusions of the paper are summarized in Section 5.

## 2 DATA AND METHODS

In this study, we draw upon the same catalogue of Gould belt molecular clouds as in Li17 (Table 1), and we again use the dust extinction maps from Dobashi (2011) to construct the MCFs. Dobashi (2011) performed fore/background reduction, which helps to identify the cloud shapes and has been adopted by us since paper I. These maps have a resolution of 1'. We selected clouds within 500 pc, which limits the distance variation between 150 and 450 pc to restrict the effect from different spatial resolutions. Still the potential effect from the factor of three difference in resolution is carefully studied (section 3.2; Appendix). The Coalsack and Cepheus nebulae are excluded because both clouds contain components that are at significantly different distances and further away than 500 pc (Coalsack: Li et al. 2013; Beuther et al. 2011; Cepheus: Schlafly et al. 2014; Zucker et al. 2019.)

　　Here we adopt the cloud-field offsets from Li17. The cloud

★ E-mail: chiyan.law@chalmers.se
† E-mail: hbli@cuhk.edu.hk

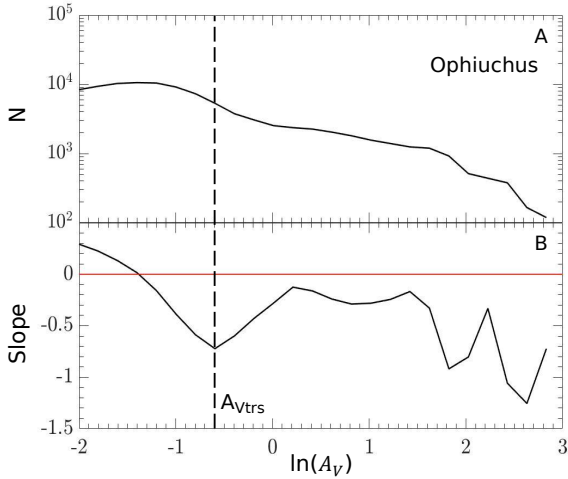

**Figure 1.** Panel A: The N-PDF of Ophiuchus derived from the dust extinction map. The bin-size of the $A_V$ is $\ln(A_V) = 0.2$. Panel B: The slope of N-PDF. The $A_{Vtrs}$ is defined by the minimum point of the first up-side-down "bell curve" like feature beyond the position which corresponds to the peak in the corresponding N-PDF (panel A).

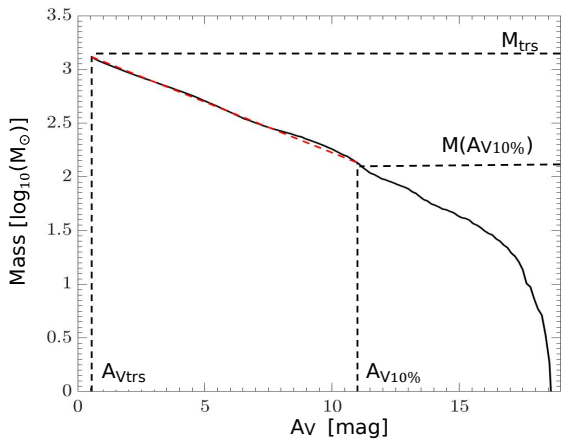

**Figure 2.** The mass cumulative function (MCF) of Ophiuchus. The starting point of the MCF is defined by the $A_{Vtrs}$. The bin size of the MCF is $A_V = 0.2$ mag. If the 10 base logarithm of cloud mass within the contour of $A_{Vtrs}$ is equal to $M_{trs}$, and the cloud mass drops to 10% of $M_{trs}$ at $A_{V10\%}$, the MCF slope (dashed line) is defined as the lower 90% of cloud mass ($log(M_{trs}) - log(0.1M_{trs}) = 1$) divided by the difference between $A_{Vtrs}$ and $A_{V10\%}$

long axis directions are determined from the auto-correlation functions of the cloud extinction maps (paper I), and the magnetic field directions are inferred from PLANCK polarimetry data (Planck Collaboration et al. 2015). The magnetic field tracers (PLANCK and optical polarimetry) are more sensitive to the bulk cloud volumes than dense cloud cores (Li17). This, in fact, is an advantage for our comparisons with star formation efficiencies (see section 4.4).

The fundamental analysis here is similar to that employed in paper II. We only use pixels with extinction higher than $A_{Vtrs}$, the "transition density" of the column density probability density functions (N-PDFs), to construct the MCFs. $A_{Vtrs}$ is defined as the density where the shape of the N-PDF changes from a log-normal to a power-law-like distribution. Simulation and observational studies have suggested that the $A_{Vtrs}$ marks the transition into the regime where gravity becomes more important than super-sonic turbulence, which is the reason for the log-normality (Vázquez-Semadeni & García 2001; Ward et al. 2014; Körtgen et al. 2019; Brunt 2015; Stutz & Kainulainen 2015; Pokhrel et al. 2016; Chen et al. 2018). We determine the $A_{Vtrs}$ based on the slope of the N-PDF (paper II). The slope of the N-PDF decreases from zero to negative from the peak of the N-PDF as the extinction increases. However, when the N-PDF deviates from the log-normal distribution to power-law at $A_{Vtrs}$, the slope of the N-PDF will increase and result in a "bell curve" feature (Figure 1 panel B). The $A_{Vtrs}$ is defined by the minimum point of the first "bell curve" feature beyond the position corresponding to the peak of the N-PDF. The essence of this method is summarized in Figure 1. Alternatively, Veltchev et al. (2019) presented a robust method to extract the $A_{Vtrs}$ based on the mathematical method "BPLFIT" (Virkar & Clauset 2014).

MCF is the amount of cloud mass that is above a given dust extinction, $\tilde{A}_V$, which is computed by the following equation (Heiderman et al. 2010):

$$MCF(\tilde{A}_V) = \mu \, m_H C \sum_{\text{pixels} (A_V > \tilde{A}_V)} A_V \times A_{\text{pixel}} \qquad (1)$$

where $\mu = 1.37$ is the mean molecular weight, $m_H$ is the mass of hydrogen, $A_{\text{pixel}} = [D(\text{cm}) \times R(rad)]^2$ is the area of one pixel, $D$ is the distance of the cloud in cm, $R$ is the pixel size in radian, and $C = 1.37 \times 10^{21} \text{ cm}^{-2} \text{ mag}^{-1}$ is the conversion factor between visual extinctions and column density (Heiderman et al. 2010). We illustrate the MCF of Ophiuchus as an example in Figure 2. Note that the MCF is log-linear ($A_V$ is binned in linear space), while the N-PDF in Figure 1 is log-log ($A_V$ is binned in log-space).

We define $M_{trs} = MCF(A_{Vtrs})$, i.e., the total mass above $A_{Vtrs}$, and $A_{V10\%}$ is defined as the extinction above which the cloud mass is 10% of $M_{trs}$. We exclude the highest 10% of $M_{trs}$ in order to distance our coverage from the cloud core regions, where mass distributions are more likely to be affected by stellar feedbacks and the scales are more likely to be beyond PLANCK resolution.

The MCF slope is defined by the following equation.

$$MCF \text{ slope} = \left| \frac{log(M_{trs}) - log(0.1M_{trs})}{A_{Vtrs} - A_{V10\%}} \right| = \frac{1}{|A_{Vtrs} - A_{V10\%}|} \qquad (2)$$

Equation 2 can also be understood as the inverse of the span of extinction that encloses the lower 90% of $M_{trs}$. The steeper the MCF slope, the smaller the extension, i.e., the less the gas density can be enhanced from $A_{Vtrs}$

## 3  RESULTS AND ANALYSIS

### 3.1  The main result

Figure 3 summarises the MCFs of all clouds, colour coded by the cloud-field offsets. Each MCF is normalized by the corresponding $M_{trs}$. We notice that molecular clouds with smaller cloud-field offsets tend to have shallower MCFs. In Figure 4, we plot the MCF slopes against cloud-field offsets and observe a positive correlation. In other words, molecular clouds with larger cloud-field offsets show steeper slopes. The horizontal error bars indicate the interquartile ranges (IQR) of the field directions of each cloud (Li17). The vertical error bars designate the uncertainties of the MCF slopes due to the MCF bin width, spatial resolution and the $A_{Vtrs}$ uncertainty (discussed in the next section).

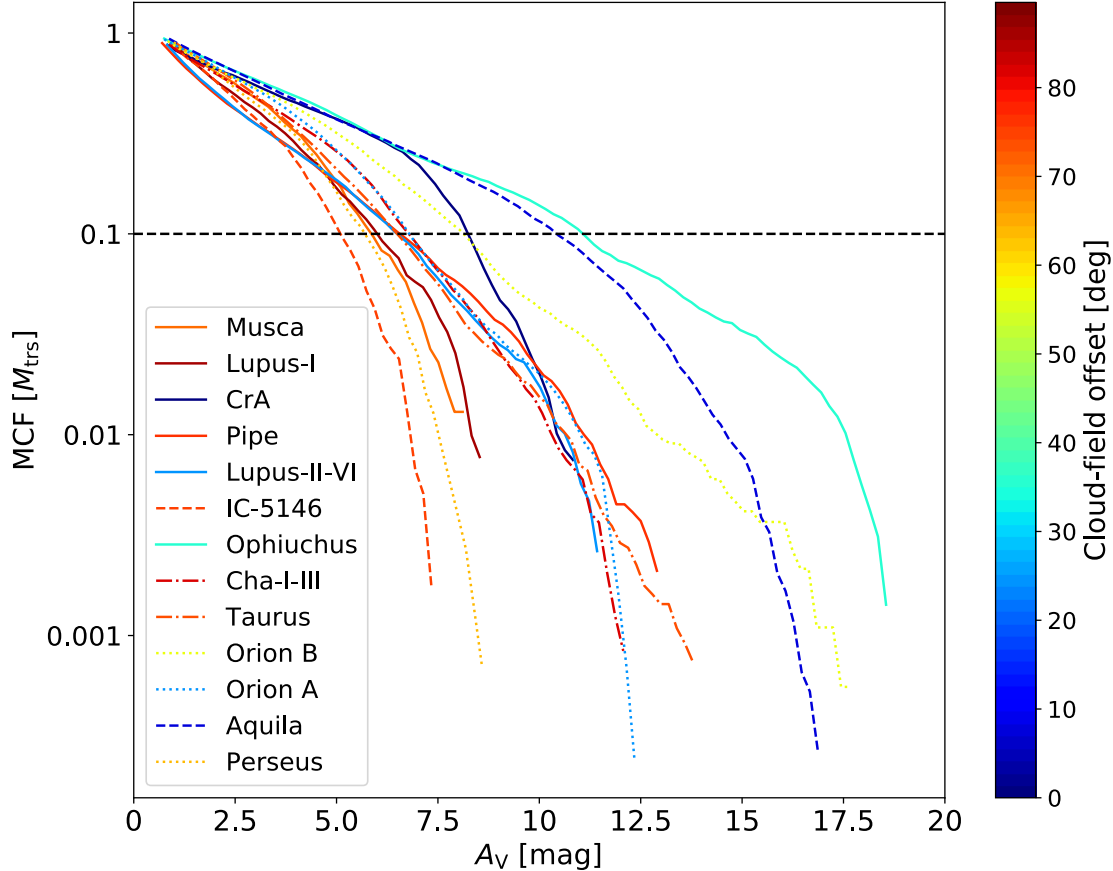

**Figure 3.** MCFs of all the Gould Belt clouds studied in this work normalized by $M_{trs}$ (Table 1). The MCFs are colour coded by the cloud-field offsets. The starting point of each MCF is defined by the corresponding $A_{Vtrs}$ of the cloud. The dashed line indicates 10% of $M_{trs}$. We notice that Perseus and Orion A/B (dotted lines) have more extensive line-of-sight scales (section 4.1) and have higher discrepancies between the magnetic field directions inferred from optical and PLANCK polarimetry data (section 4.2).

Here we study the degree of significance of the correlation in Figure 4. Due to the small sample size, we use the non-parametric permutation test. The alternative hypothesis ($H_1$) and the null hypothesis ($H_0$) are defined as follows:

$H_1 : \mu_{small} < \mu_{large}$

$H_0 : \mu_{small} = \mu_{large}$

where $\mu_{small}$ is the average MCF slope of the molecular clouds with cloud-field offsets < 45° and $\mu_{large}$ is the average MCF slope of the molecular clouds with cloud-field offsets > 45°. Then we obtain $T_{obv} = \mu_{small} - \mu_{large}$ as the difference between the two population means. The numbers of clouds with large and small cloud-field offsets are 8 and 5, respectively. We regroup the clouds into groups with sizes 5 and 8; there are $^{13}C_5 = 1287$ combinations. For each one of the combinations, we obtain the difference between the average slopes of the smaller group ($\mu_5$) and the larger group ($\mu_8$), which we denote as $T = \mu_5 - \mu_8$. The statistical significance can be quantified by the frequency of obtaining $T \leq T_{obs}$ (*p*-value). The p-value is 0.01, which means that the likelihood of obtaining $T_{obs}$ by chance in a random selection is only 1%. Hence, it is very unlikely that there is no correlation between the MCF slopes and cloud-field offsets. In short, the results of the permutation test suggests a positive correlation between MCF slopes and the cloud-field offsets.

### 3.2 The MCF slope uncertainties

Here and in the Appendixes we discuss the potential uncertainties in the MCF slopes arising from (i) extinction measurement uncertainty; (ii) line of sight (LOS) contamination; (iii) MCF bin width; (iv) spatial resolution; (v) $A_{Vtrs}$ uncertainty, and (vi) $A_V$ range adopted for the MCF slope measurements. First of all, none of the above factors should correlate with the cloud-field offsets and thus should not "create" the conclusion in the previous section. The goal of the analyses here is to study how the slope uncertainties will be reflected in the p-value. The extinction measurement uncertainty should not systematically bias particular extinction values or particular regions, and therefore should not bias the N-PDF power-law slope either. The effects on MCF ought to be even smaller, due to the summation of a large number of pixels. Furthermore, the $A_{Vtrs}$ are significantly greater than the measurement uncertainties, so shifting the MCFs by the uncertainties should not significantly affect the MCF slopes.

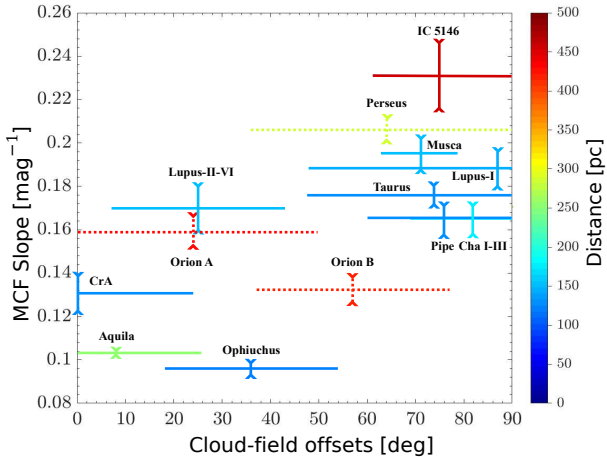

**Figure 4.** MCF slopes vs cloud-field offsets. Each symbol is colour coded by the cloud distance. The cloud-field offsets and the interquartile ranges (horizontal error bars) are adopted from Li17. The MCF uncertainties (vertical error bars) are defined based on the analyses in the Appendix (also see section 3.2). Perseus and Orion A/B (dotted lines) have more extensive line-of-sight scales (see section 4.1).

Although the LOS contamination is hard to quantify, the effects from the small-scale LOS contamination should be similar to that from the measurement uncertainty; and Dobashi (2011) had removed the large-scale (above 2°) LOS contamination. Lombardi et al. (2015) studied the bias of N-PDF due to the correction of fore/background contamination, and concluded that the bias on the power-law slope is minor. Furthermore, Ossenkopf-Okada et al. (2016) performed a detailed analysis of the effects of observational noises and fore/background contamination on the N-PDF. Their study concluded that contamination had only a marginal influence on the N-PDF power-law slope.

The discussion so far has assumed 0.2 mag (the measurement uncertainty of extinction) as the bin size for the MCF. We experimented on bin widths varying from 0.05 to 1 mag (see the Appendix for a summary). Furthermore, the differences between the spatial resolutions of the sampled clouds can be as large as a factor of three. Thus we also studied the effect from resolutions by binning the pixels to 2' and 4', i.e. factors of two and four of the original angular resolution. Also discuss in the Appendix the effects from the $A_{Vtrs}$ uncertainty and from the $A_V$ ranges selected for the MCF slope measurement

We conclude that both the MCF binning and spatial resolution have similar impacts on the slope uncertainty. On the other hand, the slope uncertainties due to $A_{Vtrs}$ uncertainty is relatively minor. Nevertheless, we estimated the total uncertainties by summing all three uncertainties quadratically, and we present them as the vertical error bars in Figure 4&5 .To study how the MCF slopes uncertainty will affect the p-value, we bootstrapped the slopes 10000 times. For every iteration the MCF slope of each cloud was randomly selected from the range defined by the slope error bar, and a corresponding p-value was calculated following the permutation test in section 3.1. The smallest interval which contains 95% of the p-value distribution (the 95% data interval) is [0.005, 0.018].

## 4 DISCUSSION

### 4.1 The effect of cloud LOS scales

Goodman et al. (1990) found that the Perseus cloud contains two structures located at different distances along the LOS direction. Recent studies by Schlafly et al. (2014) and Zucker et al. (2019) calculated accurate distances to molecular clouds based on stellar photometry and found that Perseus and Orion were more extended ("thicker") along the LOS (∼ 30pc and ∼ 44pc respectively). We examined whether the trend in Figure 4 still holds without the two special clouds, and found that the corresponding 95% data interval of the *p*−value distribution is [0.005, 0.019].

In paper II, we were not aware that Orion A/B was similar to Perseus, and excluded Perseus from the statistics as the single special case. Therefore, we applied the bootstrapping statistical test (section 3.2) to the linear mass ratio with and without Orion and Perseus. The 95% data intervals of the *p*−value distributions respectively are [0.0001, 0.003] and [0.0006, 0.018].

### 4.2 The effect of magnetic field tracers

While field directions inferred from starlight and submm polarimetry are largely in agreement, Gu & Li (2019) found discrepancies in Orion A/B and Perseus, which happen to be the thicker clouds. Hence, it is natural to study whether the trend in Figure 4 is sensitive to the magnetic field tracers. Figure 5 is similar to Figure 4, but the cloud-field offsets are computed based on starlight polarimetry from Gu & Li (2019). We perform the same statistical analyses as described in section 3.2 for the cases with and without Orion and Perseus. The 95% data intervals of the *p*−values distribution are [0.024, 0.053] and [0.008, 0.032] respectively. The results of the bootstrapping test performed in this section and section 3 are summarised in Table 2, and the *p*−value distributions are summarised in Figure 6. We notice that a positive correlation is significant when excluding the two thicker clouds, no matter which magnetic field tracer is used. When including them, only the submm tracer shows a significant correlation but not for the optical tracer.

We also repeat the bootstrapping statistical analysis on the linear mass ratio in paper II with cloud-field offsets inferred from optical polarimetry data. The 95% data intervals of the *p*−values distribution are [0.093, 0.24] for all clouds, and [0.002, 0.011] when Orion and Perseus are excluded. Thus, for the linear mass distribution, the correlations are robust for both field tracers if the thick clouds are excluded. The correlations based on PLANCK data are always significant, with or without the thicker clouds. The results are also summarised in Table 2.

Why the two field tracers disagree in the thicker clouds? An apparent link between LOS scales and the inferred magnetic field directions is that sub-millimetre data is accumulated from the entire LOS, while optical data samples only the stellar foreground. The thicker the cloud in the LOS direction, the larger the discrepancy between the LOS scales traced by sub-mm and optical data. While the former grows with the cloud thickness, the latter does not, due to the decrease of observable stars as the distance increases.

**Table 1.** General information of the 13 Gould belt molecular clouds selected for this study.

| Cloud | Distance (pc) | Long axes P.A. (°) | Magnetic field P.A.[†] (°) PLANCK | Starlight | MCF slope[★] (mag⁻¹) | $M_{trs}$ (M$_\odot$) |
|---|---|---|---|---|---|---|
| Lupus-I | $150 \pm 20^a$ | −1 | $86 \pm 37$ | $-82 \pm 13$ | $0.188 \pm 0.008$ | 224 |
| Cha-I-III | $193 \pm 39^{a\#}$ | 22 | $-76 \pm 13$ | N/A[*] | $0.165 \pm 0.006$ | 1376 |
| Pipe | $130 \pm 18^b$ | −45 | $59 \pm 16$ | $49 \pm 13$ | $0.165 \pm 0.006$ | 676 |
| IC5146 | $460 \pm 80^d$ | −38 | $67 \pm 14$ | $64 \pm 16$ | $0.231 \pm 0.015$ | 1158 |
| Taurus | $135 \pm 20^c$ | 75 | $1 \pm 27$ | $0 \pm 18$ | $0.176 \pm 0.004$ | 1785 |
| Musca | $160 \pm 25^a$ | 27 | $-82 \pm 8$ | N/A[*] | $0.195 \pm 0.007$ | 125 |
| Perseus | $294 \pm 15^e$ | 32 | $-84 \pm 28$ | $59 \pm 35$ | $0.206 \pm 0.005$ | 2727 |
| Orion B | $423 \pm 21^e$ | −30 | $-87 \pm 20$ | $7 \pm 18$ | $0.132 \pm 0.005$ | 7274 |
| Orion A | $432 \pm 22^e$ | 83 | $59 \pm 25$ | $7 \pm 20$ | $0.159 \pm 0.006$ | 12582 |
| Ophiuchus | $125 \pm 18^c$ | -45 | $-81 \pm 18$ | $-65 \pm 25$ | $0.096 \pm 0.002$ | 1305 |
| Lupus-II-VI | $160 \pm 40^{a\wedge}$ | −73 | $82 \pm 18$ | $81 \pm 11$ | $0.170 \pm 0.010$ | 743 |
| Aquila | $260 \pm 10^c$ | −75 | $-67 \pm 17$ | $-45 \pm 10$ | $0.103 \pm 0.001$ | 11005 |
| CrA | $151 \pm 10^e$ | −26 | $-26 \pm 24$ | $-27 \pm 32$ | $0.131 \pm 0.008$ | 256 |

[a]Heiderman et al. (2010). [b]Lada et al. (2010). [c]Schlafly et al. (2014). [d]Arzoumanian et al. (2011). [e]Zucker et al. (2019). [#]The distance to Chameleon is calculated by taking the average of the distances to Cha I, II & III. [∧]The distance to Lupus-II-VI is calculated by taking the average of the distances to Lupus II, III, IV, V & VI. [†]Mean directions of the magnetic field inferred from PLANCK (Li 17) and starlight polarimetry data (Gu & Li 2019). The interquartile ranges (IQRs) of the magnetic field directions are given after the mean values. [*]Musca was combined with Cha-I-III as a single structure in Gu & Li (2019). [★]The MCF slope uncertainties are obtained by summing all three uncertainties discussed in Appendix A quadratically.

**Table 2.** Permutation tests for the statistical significance (*p*-value) of the correlation between the cloud-field offset and, respectively, MCF slope (this study), and evenness of the linear mass distribution (Paper II).

| | *p*-value (95% data interval) | | | |
|---|---|---|---|---|
| | MCF Slope | | Linear mass distribution | |
| | PLANCK | Optical | PLANCK | Optical |
| All clouds | [0.005, 0.018] | [0.024, 0.053] | [0.0001, 0.003] | [0.094, 0.24] |
| Without Perseus & Orion | [0.005, 0.019] | [0.008, 0.032] | [0.0006, 0.018] | [0.002, 0.011] |

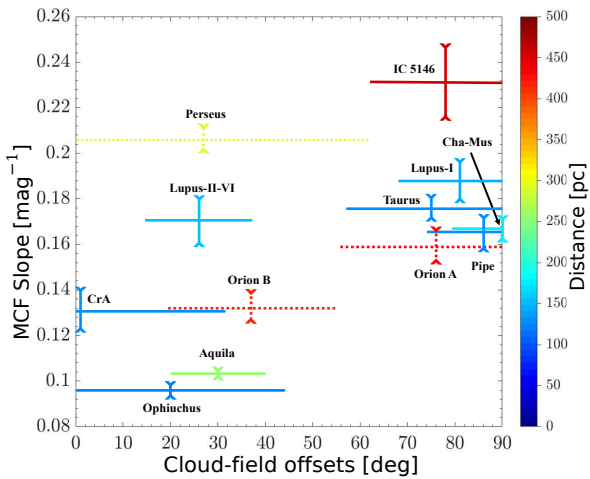

**Figure 5.** Similar to figure 4,but the magnetic field directions and error bars (IQRs) are obtained from starlight polarimetry data (Gu & Li 2019). Note that the dotted clouds (Perseus and Orion A/B) change more significantly in inferred B-field orientations from Figure 4 (see section 4.2). Musca was combined with Cha-I-III as a single structure in Gu & Li (2019), and is designated as 'Cha-Mus'.

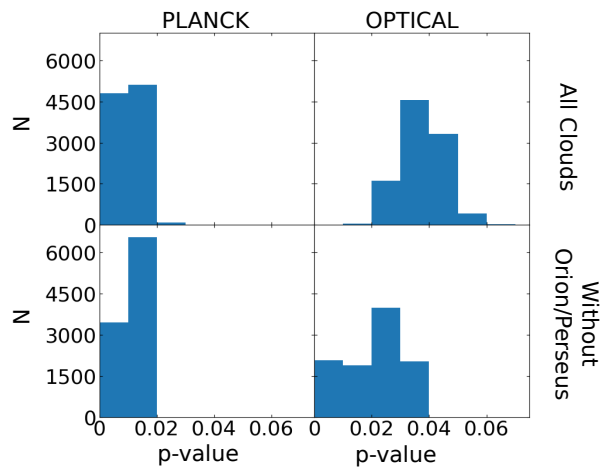

**Figure 6.** *p*−value distributions of the bootstrapping tests on PLANCK (left column) and optical (right column) data with (upper row) or without (lower row) Perseus/Orion. All p-values are less than 5% except for optical tracer with all clouds.

### 4.3 Global vs local cloud properties

We found that that local mass-to-flux ratios (Crutcher et al. 2010) and local Alfven Mach numbers (Zhang et al. 2019) both increase with cloud densities. Also, despite the global cloud-field offset bi-modality (paper I), the local structure-field offsets always tend to increase with the densities (Planck Collaboration et al. 2016). It would therefore be unreasonable to expect the field-regulation we revealed at cloud scales to remain unchanged in high density regions. We are presently analysing how the correlation varies towards higher densities (Law et al. in prep). The Herschel data will then be used as the density tracer.

However, we emphasize that the consequences of field-regulation at the cloud scales have already affected star formation through regulating gas concentration, i.e. globally parallel clouds in general result in a higher proportion of high-density gas. This effect remains for the global star formation efficiencies (SFE) of a cloud, regardless of how the importance of B-fields varies with densities. Moreover, the SFE studies (Heiderman et al. 2010; Lada et al. 2010) depend on YSO populations, which are usually distributed over the entire cloud volume instead of occurring only in high-density regions. For example, Heiderman et al. (2010) used $A_{Vtrs} = 2$ mag as the cloud threshold to estimate cloud masses and count YSOs. To study how field orientations can affect SFE, global cloud-field offsets should therefore be used, as we have done in this series. Concentrating merely on dense regions and local structure-field offsets within them (e.g. Soler 2019) will fail to capture the process whereby cloud-scale fields regulate gas concentration and thus star formation

### 4.4    Comparison with papers I and II

Figure 7 presents a summary of the findings from this work and papers I & II. Molecular clouds with smaller cloud-field offsets display shallower MCF slopes and less even mass distributions. These findings imply that cloud-field parallelism helps gas accumulate towards one side of a cloud, presumably because cloud-field offsets affect the magnetic flux of the cloud. In the case of a cylindrical cloud, a parallel cloud-field offset possesses lower magnetic flux, because the cross-section of the cloud perpendicular to the B-field is smaller. Intuitively, in the parallel cases, gas can accrete freely along the cloud long axis, but this would be hindered by the Lorentz force in the perpendicular cases.

## 5    CONCLUSIONS

We here present an analysis of the slopes of the mass cumulative functions (MCFs) of 13 Gould belt molecular clouds within 500 pc. We investigate the effects of cloud-field alignments on the cloud MCF slopes, and find that molecular clouds aligned perpendicularly to the mean direction of the ambient magnetic field tend to have steeper MCF slopes,meaning that less mass resides in high extinction. Permutation and bootstrapping tests demonstrate a positive correlation between the MCF slopes and cloud-field offsets. The correlation is robust ($p < 0.05$) to the field tracers (optical and submillimeter polarimetry). Together with the correlation between the linear mass distribution and cloud-field offset (Paper II) and the bimodality of star formation efficiency versus cloud- field offset (Li17), the evidence for field-regulated cloud fragmentation is becoming ever more compelling as the number of tests accumulate.

## ACKNOWLEDGEMENTS

The authors would like to acknowledge the funding they have received from the Research Grants Council of Hong Kong: General Research Fund grants Nos. 14307118, 14307019, 14305717 and 14304616. Law, CY thanks Leung, P.K. for fruitful discussions, encouragement and help. The authors would also like to thank the anonymous referee for valuable comments and suggestions.

## DATA AVAILABILITY

The data underlying this article will be shared on reasonable request to the corresponding author.

## APPENDIX A:    EFFECTS OF DIFFERENT FACTORS ON THE MCF SLOPES

In addition to the LOS contamination and measurement uncertainty discussed in section 3.2, the uncertainties of $A_{Vtrs}$, the spatial resolution uncertainties of the extinction map, and the bin sizes of the MCF may also affect the MCF slope. We obtain the total MCF slope uncertainties (Table 1 column 6) by summing all the uncertainties of the three factors discussed below quadratically.

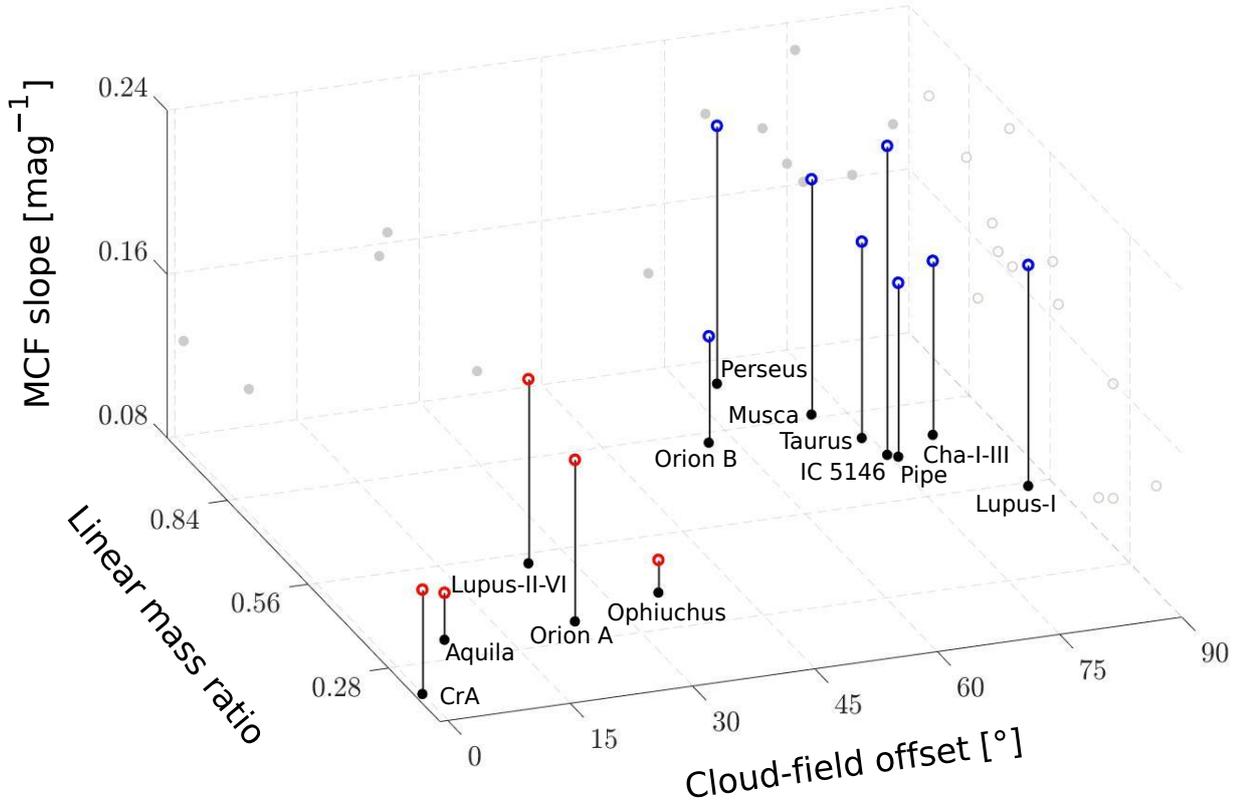

**Figure 7.** MCF slopes vs linear mass ratios vs cloud-field offsets of the 13 Gould belt clouds studied in this work and paper II. The linear mass ratios are adopted from paper II. The MCF slopes are from Table 1. The cloud-field direction offsets inferred from PLANCK polarimetry data are adopted from Li17. Most molecular clouds with cloud-field offsets smaller/larger than 45 deg have MCF slopes below/above 0.16 and mass ratios below/above 0.56.

**$A_{Vtrs}$ uncertainty**

In this study, the boundary of each cloud region is determined by the $A_{Vtrs}$, which is used to define the starting point of the MCF slope. The $A_{Vtrs}$ is determined based on the slope of cloud N-PDF (Figure 1). Here we study the effect of the N-PDF bin size on $A_{Vtrs}$ and MCF slopes by setting the bin size to $\ln(A_v) = [0.05, 0.1, 0.3, 0.4]$ (see Figure A1 for an illustration). The $A_{Vtrs}$ will vary with the bin size and result in a MCF slope dispersion for each cloud is less than 0.005 mag$^{-1}$. Care should be taken to avoid using too narrow or wide bin sizes: the statistical noise emerges if the bin is too fine, while the significant features in the N-PDF will be smoothed out if the bin is too wide (Schneider et al. 2015).

**MCF bin width**

We study the effect of bin width of MCF on its slope by sampling a range of bin sizes ($A_v = [0.05, 0.1, 0.3, 0.5, 0.7, 0.9, 1]$ mag). Figure A2 presents the MCF slopes as a function of MCF bin widths. The slope dispersion for each cloud is less than 0.01 mag$^{-1}$.

**Extinction map resolution**

To study the effects of distance uncertainty on the MCF slopes, we degrade the pixel resolution from 1' to 2' and 4' and observe how the MCF slope will change accordingly. Figure A3 summarises the results. We find that the slope dispersion for each cloud is less than 0.015 mag$^{-1}$. Therefore, the MCF bin sizes and the resolution have a similar impact on the MCF slopes.

**APPENDIX B: MCF RANGES USED TO MEASURE THE SLOPES**

To study whether the correlation in Figure 4 is sensitive to the MCF range adopted to derive the slopes, we perform a piece-wise fit of the slope with two linear functions, leaving the turning point as one parameter to fit (see Figure B1 for an illustration). We take the slope of the first linear fit (light blue line) as the MCF slope. Table B1 summarises the MCF slopes of all clouds based on a piece-wise fit. We evaluate the statistical significance of the trend by performing a permutation test (section 3.1) on the fitted MCF slopes with and without Orion A/B and Perseus. The corresponding $p$–values are summarized in Table B2. The correlation is robust for both tracers. Together with Table 2, where the MCF slopes are defined by different MCF ranges, we argue that our results are not sensitive to the method of measuring slopes.

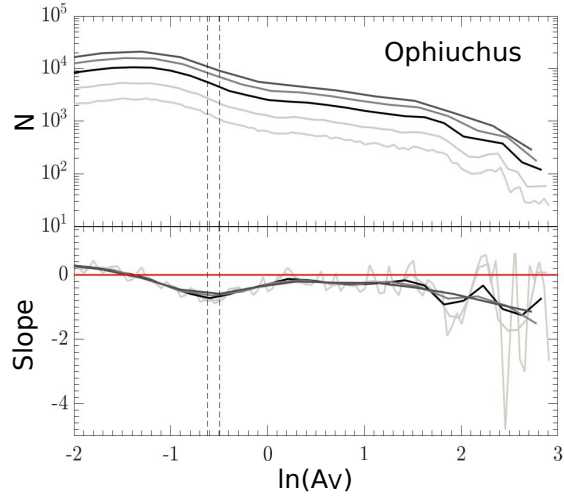

**Figure A1.** Illustration of the $A_{Vtrs}$ uncertainty due to varies N-PDF bin sizes. The black solid line shows the N-PDF with logarithmic bin size ln(Av) = 0.2. The remaining lines with decreasing grayness present the N-PDFs with logarithmic bin sizes ln(Av) = [0.4, 0.3, 0.1, 0.05]. The two vertical dashed lines indicate the upper and lower bounds of the variations in the $A_{Vtrs}$, due to the effects of bin width.

**Table B1.** Fitted low density MCF slopes with 95% confidence interval of the 13 Gould belt clouds.

| Cloud | MCF slope (mag$^{-1}$) |
|---|---|
| Lupus-I | 0.196 ± 0.00663 |
| Cha-I-III | 0.169 ± 0.0107 |
| Pipe | 0.162 ± 0.00243 |
| IC5146 | 0.236 ± 0.0192 |
| Taurus | 0.201 ± 0.00514 |
| Musca | 0.157 ± 0.0111 |
| Perseus | 0.206 ± 0.0141 |
| Orion B | 0.123 ± 0.00895 |
| Orion A | 0.197 ± 0.00669 |
| Ophiuchus | 0.104 ± 0.00211 |
| Lupus-II-VI | 0.178 ± 0.00421 |
| Aquila | 0.111 ± 0.00539 |
| CrA | 0.095 ± 0.00437 |
| Cha-Mus | 0.162 ± 0.0122 |

**Table B2.** Results of permutation tests on the correlation between cloud-field offsets and fitted MCF slopes.

| | *p*-value | |
|---|---|---|
| | PLANCK | Optical |
| All clouds | 0.043 | 0.017 |
| Without Perseus & Orion | 0.019 | 0.024 |

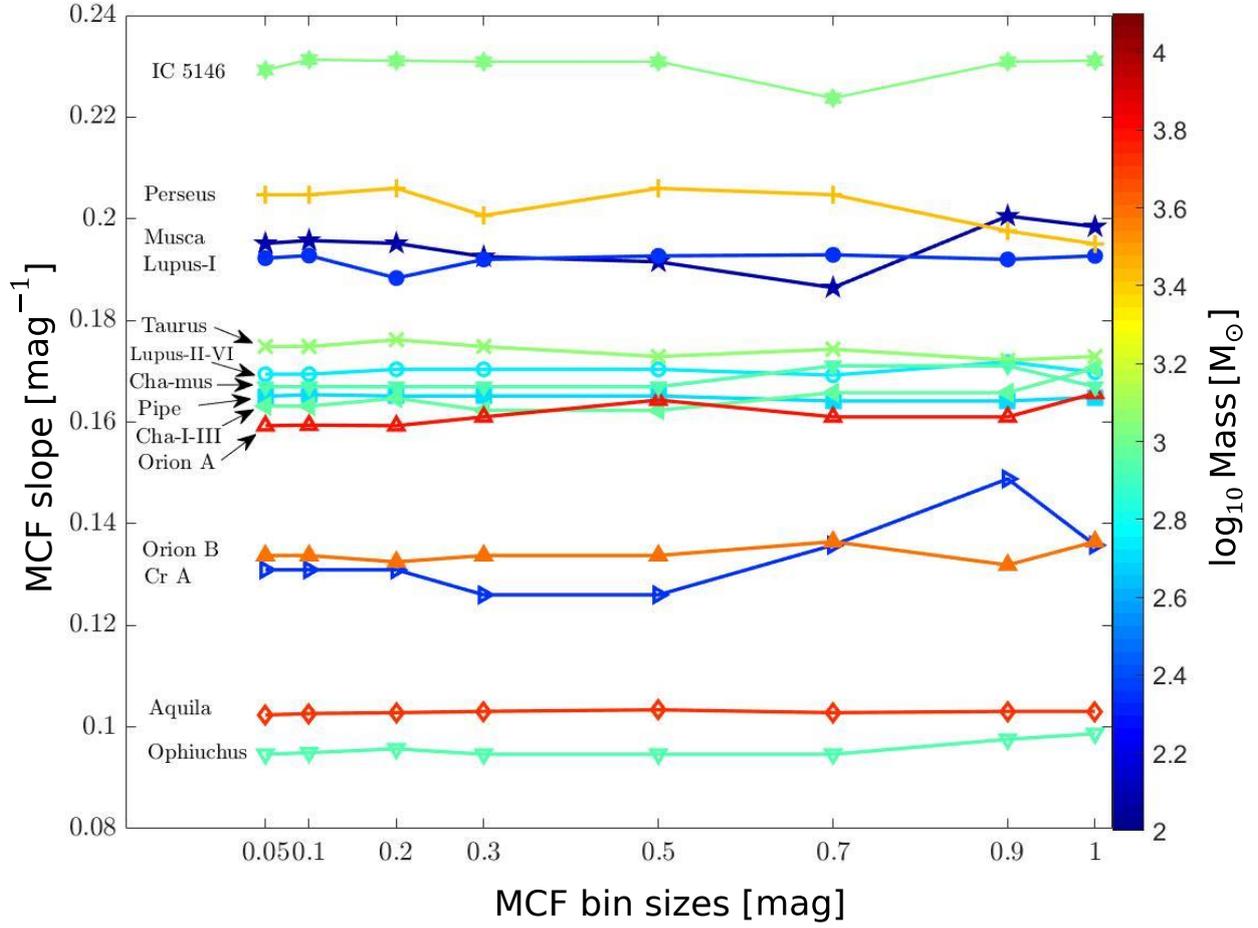

**Figure A2.** MCF slopes as a function of MCF bin sizes for all clouds in this study. The symbols are colour coded by $M_{lrs}$ (Section 2). Filled and opened symbols indicate large and small cloud-field offsets. We also notice that there is no correlation between $M_{lrs}$ and the MCF slopes.

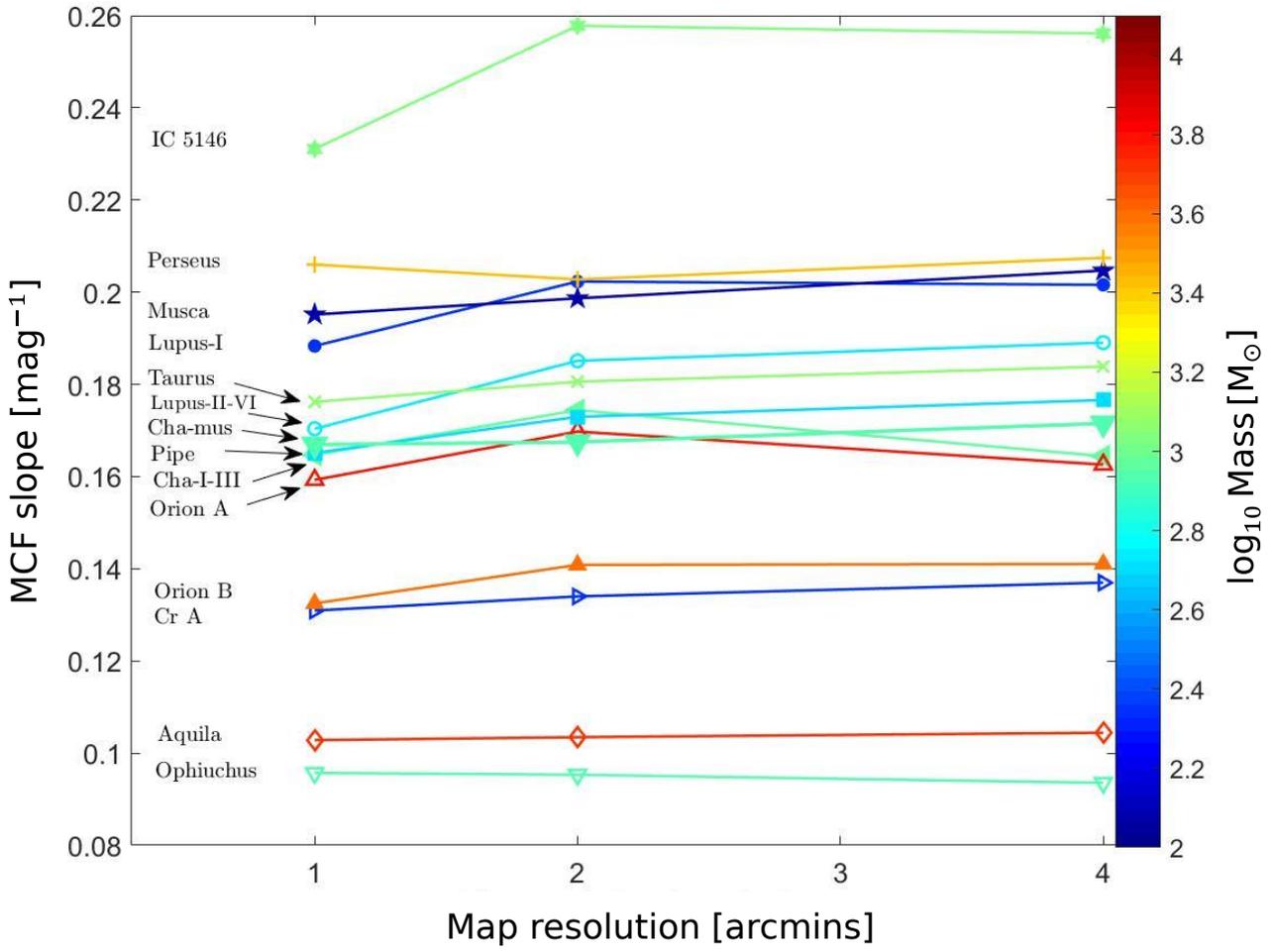

**Figure A3.** MCF slopes as a function of map resolutions for all clouds in this study. The symbols and color codes are the same as Figure A2.

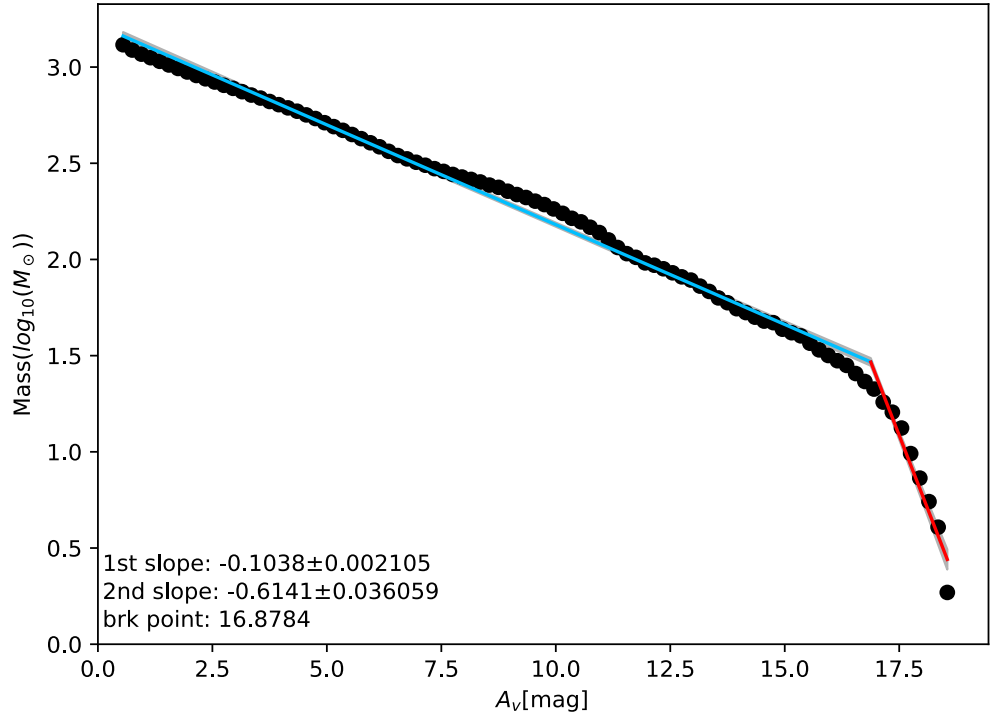

**Figure B1.** MCF of Ophiuchus overlaid with the piece-wise linear fit. We take the slope of the first linear fit (light blue line) as the MCF slope.

This paper has been typeset from a TeX/LaTeX file prepared by the author.